\documentclass[conference]{IEEEtran}
\IEEEoverridecommandlockouts
\usepackage{cite}
\usepackage{amsmath,amssymb,amsfonts}
\usepackage{algorithmic}
\usepackage{graphicx}
\usepackage{textcomp}
\usepackage{xcolor}
\usepackage{subcaption}
\usepackage{booktabs}
\usepackage{stfloats}

\def\BibTeX{{\rm B\kern-.05em{\sc i\kern-.025em b}\kern-.08em
    T\kern-.1667em\lower.7ex\hbox{E}\kern-.125emX}}
\begin{document}

\title{CONVERGE: A Multi-Agent Vision-Radio Architecture for xApps}

\author{\IEEEauthorblockN{Filipe B. Teixeira\IEEEauthorrefmark{1},
Carolina Simões, Paulo Fidalgo, Wagner Pedrosa, André Coelho, \\Manuel Ricardo, and
Luis M. Pessoa}
\IEEEauthorblockA{INESC TEC, Faculdade de Engenharia, Universidade do Porto, Portugal}
\IEEEauthorblockA{\IEEEauthorrefmark{1}e-mail: filipe.b.teixeira@inesctec.pt}% Corresponding author
\thanks{This work was supported by the CONVERGE project which has received funding under the European Union’s Horizon Europe research and innovation programme under Grant Agreement No 101094831.}  
}

\maketitle

\begin{abstract}
Telecommunications and computer vision have evolved independently. With the emergence of high-frequency wireless links operating mostly in line-of-sight, visual data can help predict the channel dynamics by detecting obstacles and help overcoming them through beamforming or handover techniques. 

This paper proposes a novel architecture for delivering real-time radio and video sensing information to O-RAN xApps through a multi-agent approach, and introduces a new video function capable of generating blockage information for xApps, enabling Integrated Sensing and Communications. Experimental results show that the delay of sensing information remains under 1\,ms and that an xApp can successfully use radio and video sensing information to control the 5G/6G RAN in real-time. 
\end{abstract}


\section{Introduction}
Integrated Sensing and Communications (ISAC) is emerging as a key 6G trend, driven by the shift towards higher frequency bands and larger antenna arrays, including Large/Reconfigurable Intelligent Surfaces (LIS/RIS). This trend is bringing communications and sensing systems closer. This convergence is expected to enhance wireless networks with environmental sensing capabilities and increase resource usage efficiency~\cite{Saa19}. The integration of RIS into wireless environments is expected to improve the performance of both communications and sensing, particularly under non-line-of-sight conditions~\cite{Liu22}. ISAC is also creating an opportunity for integrating wireless communications and Computer Vision (CV) in a promising multimodal approach, unlocking the potential for high-resolution sensing. Applications range from device localisation to device-free environment imaging/mapping, including human sensing, with immense market potential in different verticals. CV enables low computationally complex user/object tracking and environmental awareness, a capability that complements well radio communications-based sensing. While the latter is computationally more complex, it deals better with obstructions or limited lighting conditions. This intersection of domains creates an opportunity for new AI-based heterogeneous and multimodal data fusion solutions, pivotal in enabling 6G high-frequency communications. Addressing this interdisciplinary challenge requires advanced Research Infrastructures (RI) and suitable tools. CONVERGE is a pioneering vision-radio RI that bridges this gap by leveraging ISAC to facilitate a dual ``View-to-Communicate, Communicate-to-View'' approach~\cite{EuCNC24}. The xApps defined by the 5G RAN Intelligent Controller (RIC) to control in near real-time the Radio Access Network (RAN) may play a relevant role as ISAC-based controllers in CONVERGE.

This paper introduces a novel architecture, developed within the scope of CONVERGE, capable of delivering real-time radio and video sensing information to xApps through a multi-agent approach. Since xApps target enabling the near real-time control of the RAN, for example by considering applications involving beam management from base stations or RISs towards mobile devices, it is therefore essential to experimentally evaluate the timescales involved in delivering sensing information from the agents to the xApp. We perform this experimental evaluation under different dynamic constraints, such as the number of agents, message size, and message throughput, in order to assess the potential impact of the proposed architecture for real-time sensing applications. We also consider a radio signal blockage use case to demonstrate the combination of radio and video sensing on an xApp.

The main contributions of this paper are twofold: 1) a multi-agent system capable of conveying radio and video sensing information to xApps in near real-time; 2) a new video unit function capable of generating relevant video sensing/blockage messages in real time to be sent by an agent to the xApps.

The paper is structured as follows. Section II  reviews the state of the art regarding the integration of CV with the RAN, considering  the O-RAN architecture. Section III introduces the novel multi-agent architecture. Section IV details the implementation of a system based on the proposed architecture. Section V describes the validation carried out  by means of a use case testing scenario. Section VI provides the main conclusions and directions for future work.

\section{Review of the state of the art}

\begin{table}[ht]
\centering
\vspace*{4pt} 
\caption{Comparison of existing works addressing the integration of radio or video sensing into O-RAN architecture.}
\label{tab:comparison}
\setlength{\tabcolsep}{4pt} % Reduce space between columns
\begin{tabular}{p{1.0cm} p{4.5cm} p{2cm}}
\toprule
\centering\textbf{Work} & \centering\textbf{Approach} & \textbf{Enablers} \\ 
\midrule
\cite{Charan2021} & ML and video data at BS for proactive blockage prediction and seamless handover & YOLOv3, ML  \\ %\hline
\cite{Alrabeiah2020} & CV predicts mmWave beams and blockages from RGB images and sub-6\,GHz channels for proactive beam/blockage prediction using syntethic datasets & RGB images, CV, ML \\ %\hline
\cite{Bonati2022} & Framework for data-driven experimentation in O-RAN, supporting xApp development and testing for enhancing closed-loop control & OpenRAN Gym, xApps \\ %\hline
\cite{Maia2022} & gNB mounted on mobile robotic platform with real-time video-based human control through a novel On-Demand Mobility Management Function (ODMMF) & Video cameras on a mobile robotic platform, OAI, FlexRIC, OAI\\ %\hline
\cite{Queiros2024} & O-RAN with gNB carried by a mobile robotic platform. SNR is used by another xApp for positioning the gNB, in order to enhance UE link quality & SNR samples, mobile robotic platform, xApp  \\ %\hline
\cite{Kim2024,ahn2022toward} & CV for improved beam management & CV \\
\bottomrule
\end{tabular}
\end{table}

A limited number of works address the integration of CV into the O-RAN architecture. In~\cite{Charan2021}, a machine learning (ML) solution leverages visual data from video cameras at base stations (BS); by employing YOLOv3~\cite{Redmon2018}, this solution uses bimodal data to proactively predict blockages and enable seamless handovers. In~\cite{Alrabeiah2020}, CV addresses challenges in mmWave wireless systems; this approach predicts mmWave beams and blockages directly from RGB images and sub-6\,GHz channels, removing the need for explicit channel knowledge. \cite{Alrabeiah2020} conducts an evaluation using visual data to predict mmWave dynamic link blockages using two synthetic datasets generated with the ViWi framework~\cite{Alrabeiah2019}; this prediction enables wireless networks to proactively manage beam switching and hand-offs. \cite{Bonati2022} introduces OpenRAN Gym, a framework for data-driven experimentation within the OpenRAN paradigm, enhancing closed-loop control; it supports the development, training, and testing of xApps, integrating service models with RAN nodes. \cite{Maia2022} discusses the development of a gNB mounted on a mobile robotic platform; this solution offers wireless connectivity for UEs and includes a novel an On-Demand Mobility Management Function (ODMMF) that monitors radio conditions and allows for real-time human control using video cameras on-board the mobile robotic platform. \cite{Queiros2024} proposes a mobile O-RAN with a gNB deployed on a mobile robotic platform capable of autonomous positioning; it also proposes a novel Mobility Management xApp that uses Signal-to-Noise Ratio (SNR) samples to position the mobile RAN, thus enhancing UE link quality. \cite{Kim2024,ahn2022toward} use CV information to improve beam management techniques. Table~\ref{tab:comparison} summarizes the existing work on the integration of CV in the O-RAN architecture. 

\begin{figure*}[ht!]
  \centering
  \includegraphics[width=0.94\textwidth]{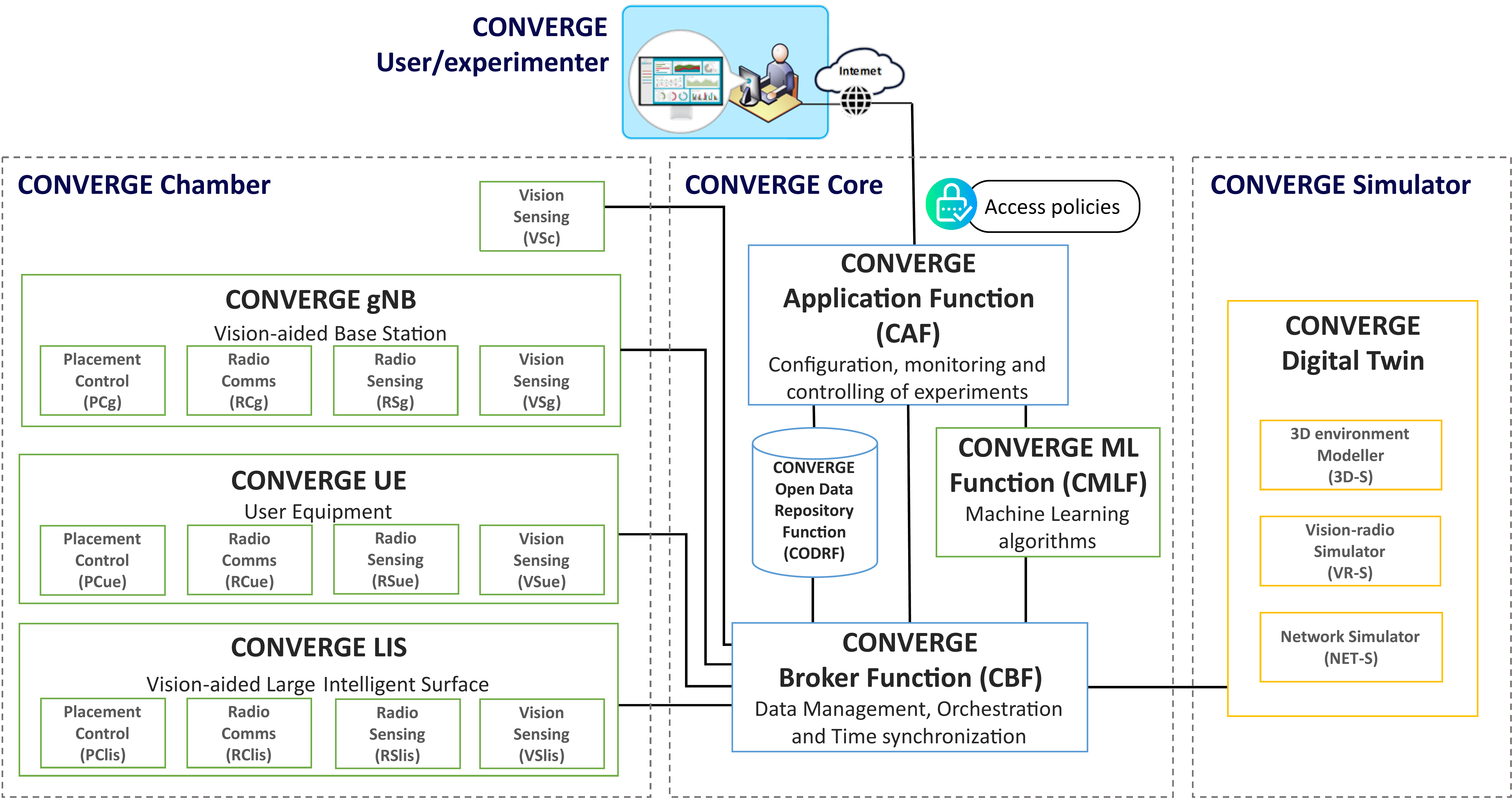}
  \caption{CONVERGE high-level architecture.}
  \label{fig:CONVERGE_high_level_architecture}
\end{figure*}

FlexRIC\cite{flexric,schmidt2021flexric} is a flexible, modular, and programmable RIC platform used in O-RAN-based 5G and beyond. It is designed to enable near-real-time dynamic control and optimization of RAN functions by implementing custom control applications known as xApps that can be tailored to specific use cases. The usage of FlexRIC has been demonstrated in~\cite{bimoflexric}, where interoperability tests between FlexRIC and O-DU for slice management and  radio resource management (RRM) were performed. FlexRIC was also used to successfully monitor and control the RAN in~\cite{ferreira2023enhancing}.

CONVERGE presents a pioneering vision-radio research infrastructure (RI)  by leveraging ISAC to facilitate a dual “View-to-Communicate, Communicate-to-View” approach. CONVERGE develops tools that are integrated in CONVERGE anechoic chambers, enabling the collection of experimental data from radio communications, radio sensing, and vision sensing. CONVERGE is aligned with the ESFRI SLICES-RI~\cite{Fdi2022} and will serve as an RI that will provide the scientific community with open datasets of both experimental and simulated data, enabling research in 6G and beyond addressing various verticals, including telecommunications, automotive, manufacturing, media, and health. CONVERGE is based on OpenAirInterface (OAI)~\cite{oai} O-RAN and uses FlexRIC for its near-RT RIC.

\begin{figure*}[b]
  \centering
  \includegraphics[width=1.0\textwidth]{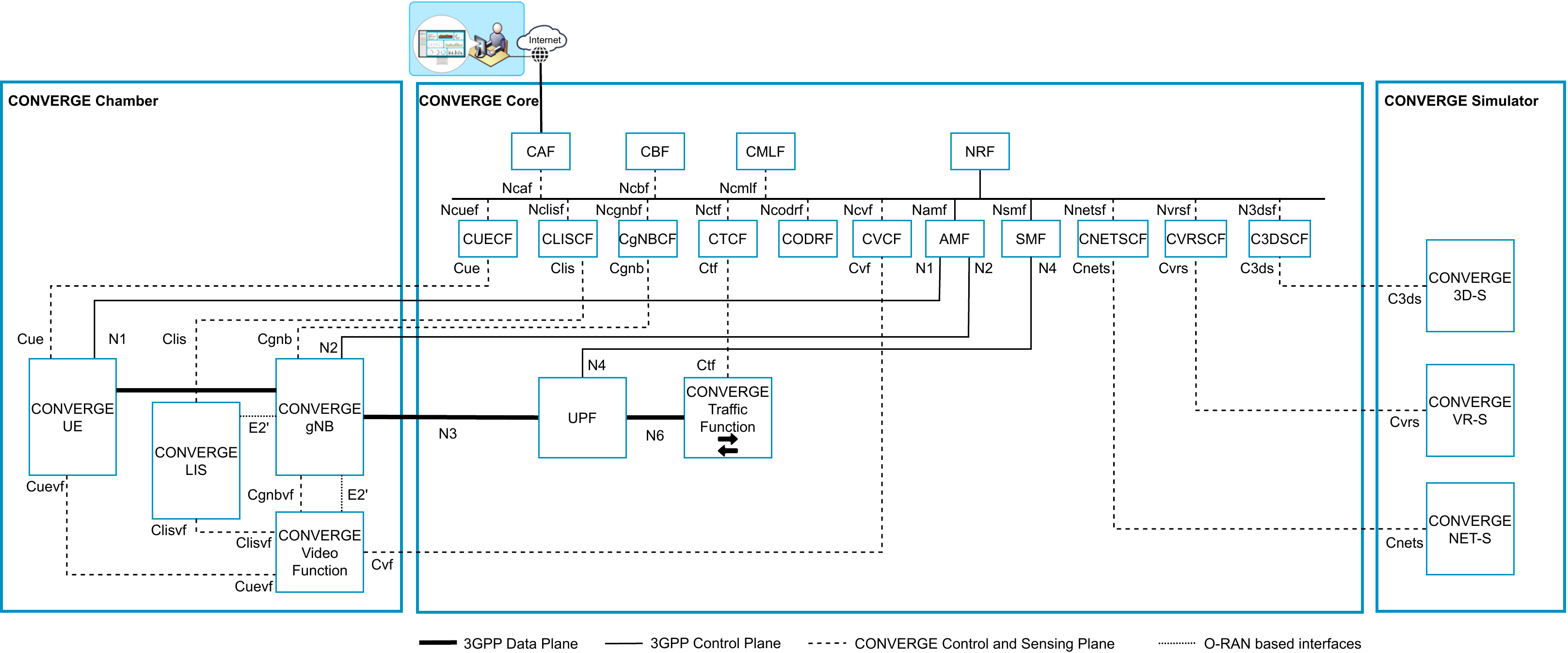}
  \caption{CONVERGE service-oriented architecture.}
  \label{fig:CONVERGE_service_oriented_architecture}
\end{figure*}

\section{Proposed Multi-Agent Architecture}

The multi-agent approach is based on the CONVERGE vision-radio RI architecture. Fig.~\ref{fig:CONVERGE_high_level_architecture} introduces the key building blocks of the CONVERGE system and Fig.~\ref{fig:CONVERGE_service_oriented_architecture} decomposes these blocks, presents their control as virtual network functions, and identifies the interfaces. 

\subsection{CONVERGE High-level architecture}
The CONVERGE architecture, depicted in Fig.~\ref{fig:CONVERGE_high_level_architecture}, comprises three main components: the CONVERGE Chamber, which facilitates physical experiments with gNB, UE, and LIS equipment across various research infrastructures; the CONVERGE Simulator, which supports scenario planning and testing via a Digital Twin; and the CONVERGE Core, which manages experiments, data, and ML models. This integrated setup enables both physical and simulated studies, allowing external users to perform real-time physical or offline virtual experiments utilizing the chamber, simulator, ML tools, and datasets for advanced research. 

The CONVERGE Chamber is equipped with a gNB, a UE, and a LIS, each featuring controls for placement (PCg, PCue, PClis), radio communications (RCg, RCue, RClis), radio sensing (RSg, RSue, RSlis), and video sensing (VSg, VSue, VSlis). The chamber also has a video sensing capability (VSc). The CONVERGE Core oversees operations, enabling user interaction for experiment setup, monitoring, and control. It includes the CONVERGE Application Function (CAF) for user interface, the CONVERGE Broker Function (CBF) for session orchestration, the CONVERGE Open Data Repository Function (CODRF) for data storage, and the CONVERGE Machine Learning Function (CMLF) for accessing ML tools. The CONVERGE Simulator, with its 3D Environment Modeller (3D-S), Vision Radio Simulator (VR-S), and Network Simulator (NET-S), creates a Digital Twin of the physical environment.

\subsection{CONVERGE Service-oriented architecture}
The CONVERGE architecture adopts a service-oriented design aligned with the 5G Core network (Fig.~\ref{fig:CONVERGE_service_oriented_architecture}), based on 3GPP standards \cite{3gpp}, leveraging 5G Core functionalities such as the Access and Mobility Management Function (AMF), the Session Management Function (SMF), and the User Plane Function (UPF), necessary for UE and gNB operations, while using the 5G New Radio (NR) specifications and facilitating the integration of new functions for CONVERGE operation. CONVERGE adds functions to the control and sensing plane to perform a variety of tasks. The CONVERGE Video Function (CVF) receives and processes the video feeds from the cameras in the CONVERGE Chamber. The control of the CONVERGE Chamber equipment and CONVERGE Simulator tools is performed using different functions (CUECF, CLISCF, CgNBCF, CTCF, CODRF, C3DSCF, CVRSCF, and CNETSCF), based on REpresentational State Transfer (REST) web services. More details on these functions can be found in~\cite{CONVERGE}.

\begin{figure}[t!]
  \centering
  \includegraphics[width=0.50\textwidth]{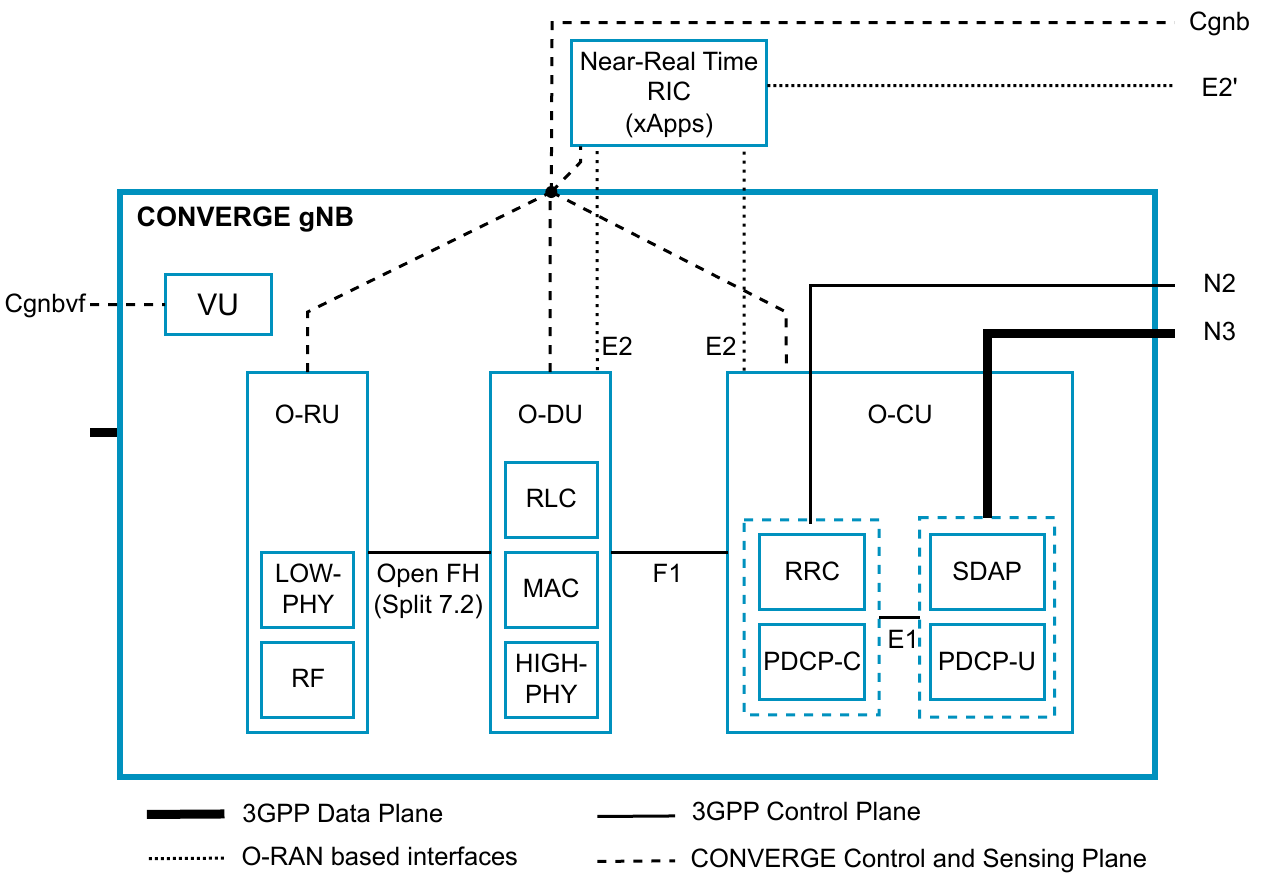}
  \caption{CONVERGE gNB architecture.}
  \label{fig:CONVERGE_gnb_architecture}
\end{figure}

\subsection{gNB architecture}
The vision-aided gNB is based on the O-RAN architecture, as shown in Fig.~\ref{fig:CONVERGE_gnb_architecture}. The gNB integrates the key components of O-RAN: the Radio Unit (O-RU), Distributed Unit (O-DU), and Centralized Unit (O-CU), following the 3GPP F1 interface and functional split 7.2~\cite{3gpp}. Vision-sensing capabilities are added through a video unit (VU). The CONVERGE gNB interfaces with the CONVERGE Core through a web service-based interface for remote access and control (CgNB). The gNB also includes the Near-Real Time RIC for the development and validation of xApps, which will be used to improve network operations by using both radio sensing and vision sensing data.

\subsection{Multi-agent architecture}
\begin{figure}[ht!]
  \centering
  \includegraphics[width=0.45\textwidth]{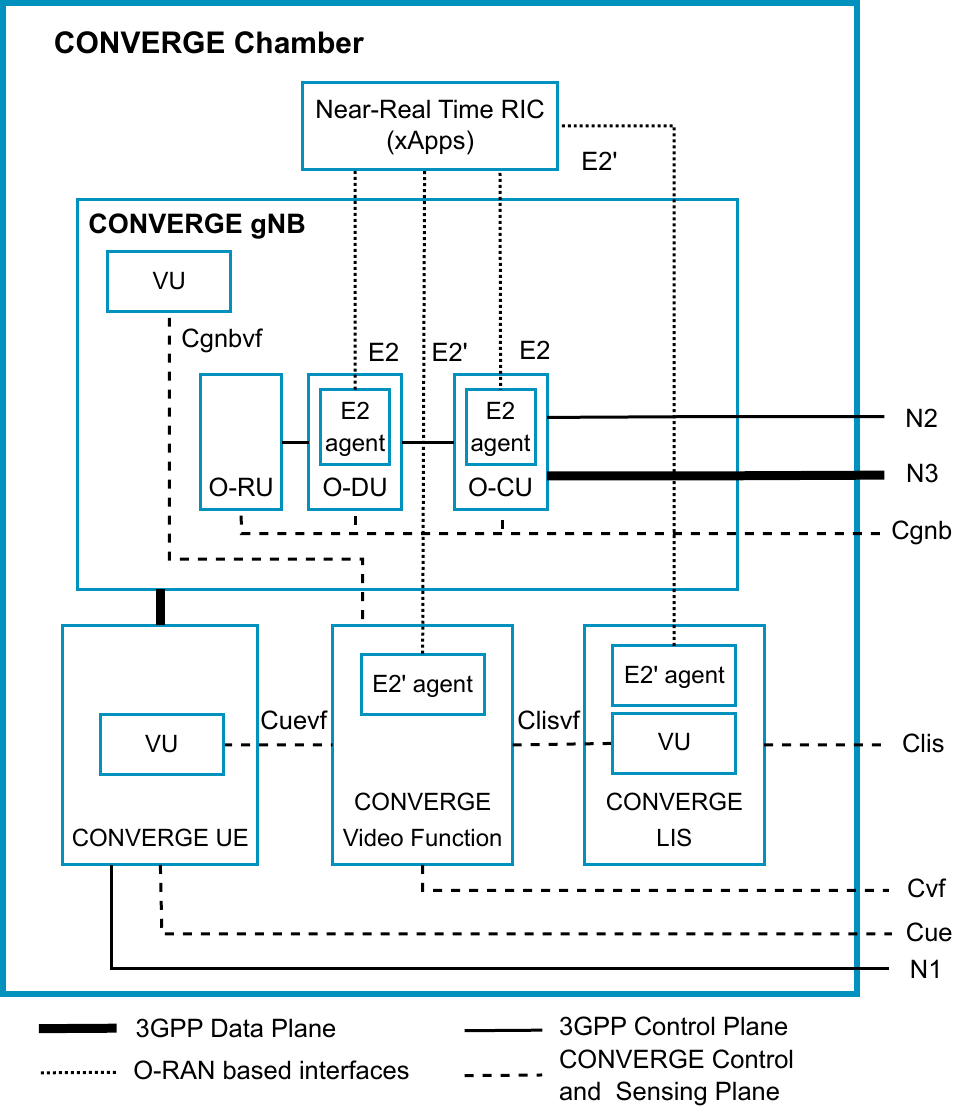}
  \caption{Detail of the multi-agent architecture.}
  \label{fig:CONVERGE_chamber_detail}
\end{figure}

The CONVERGE RIC is supported by the multi-agent architecture shown in Fig.~\ref{fig:CONVERGE_chamber_detail}. The RIC receives radio sensing data from the E2 agents on O-DU and O-CU through the O-RAN E2 interface. We extend this functionality through a new set of E2' agents that will enable the RIC to gather video and sensing data from CVF and CONVERGE LIS, respectively. This novel E2' interface has a similar operation as the E2 interface but enables the connection to agents outside the RAN.

\section{Implementation}
This section describes the implementation of the proposed architecture which includes a testbed for sensing data delay measurements and a novel vision-aided xApp integrated in a standalone 5G network. We have considered OpenAirInterface (OAI)\cite{oai} and Mosaic5G’s FlexRIC~\cite{flexric} for the Near-RT RIC due to its modular design.

\subsection{Multi-agent system}
In order to evaluate the delay and the capability of receiving and processing data from multiple agents, a system was built composed by three virtual machines: one implementing the FlexRIC
and two other machines implementing two E2' agents. The machines were based on i7-8700 CPU, with 1 vCPU and 2\,GB of RAM, running Ubuntu 24.04 and FlexRIC v1.0.0 and synchronized through Precision Time Protocol (PTP). A custom xApp was built to accurately record delays. The E2 agents were enhanced by including a timestamp in every message, enabling precise latency comparisons on the xApp. Additionally, to control message sizes, an array of 8-bit unsigned integers was integrated into the existing message structure. Since the original setup supported a minimum interval of 1\,ms between exchanged messages, adjustments were made to the timing mechanism to allow for shorter intervals.

The CVF and FlexRIC exchange data via the new E2' interface, which is tailored to carry information, such as obstacle detection, through a novel set of messages generated by CVF, detailed below. Abstract Syntax Notation One (ASN.1) for encoding and decoding these messages was used. ASN1Tools~\cite{asn1tools} was chosen for the CVF because it offers a simple API for handling ASN.1 data structures. The connection between FlexRIC and xApp is performed through the E42 interface, which offers similar functionality as E2, including: 1) Setup Request, 2) Setup Response, 3) Subscription Request, 4) Subscription Delete Request, and 5) E42 RIC Control Request.

In the CVF, OpenCV \cite{opencv} is used to capture video frames and detect ArUco markers~\cite{Garrido-Jurado2014}. ArUco markers help identify UEs without the need to train the YOLO model to recognize such objects. Ultralytics YOLO~\cite{ultralytics_docs} is employed for real-time object detection, known for its speed and accuracy. YOLO maintains continuous track of object identification and predicts future positions of obstacles across frames. OpenCV and Ultralytics YOLO together provide detection, tracking, and message exchange functionalities. This integrated approach allows the CVF to monitor and report obstacles. The YOLOv8n version was chosen for its reduced parameter count, making it less resource-intensive and well-suited to our solution. This accelerates processing times and reduces CPU load, enhancing system efficiency and responsiveness. A pre-trained model was sufficient for our solution, reducing the CPU requirements. A full HD camera, acting as the VU, was used at 30 frames per second (fps).

\subsection{\label{sec:cvf-messages}CONVERGE Video Function messages}
CVF processes video frames to detect and track obstacles, transmitting this information to the xApp through FlexRIC. This system interfaces with O-RAN xApps connected to it. The CVF generates three types of messages: \emph{Prior Blockage}, \emph{Blockage}, and \emph{Post Blockage} messages.

\emph{Prior Blockage} messages are generated when an obstacle is predicted to obstruct the line-of-sight (LoS) between the gNB and the UE, based on the obstacle's current trajectory. To predict a blockage, CVF gathers the tracking history of the obstacle across several frames using YOLO and BoT-SORT\cite{aharon2022bot} algorithms. By establishing a tracking history and assuming a constant velocity for the obstacle’s movement, the CVF calculates the obstacle’s velocity to predict its future positions in upcoming video frames. This enables the system to assess potential interruptions of the LoS. \emph{Blockage} messages are generated when a blockage of the LoS by an obstacle is detected. The system verifies whether the current position of the obstacle matches the last known position where the ArUco marker was detected and associated with the UE. If there is an overlap between the obstacle’s most recent position and the UE's last recorded position, a \emph{Blockage} message is generated. \emph{Post Blockage} messages are generated when an obstacle no longer blocks the UE, based on the updated list of the status for each tracked obstacle. \emph{Post Blockage} messages inform that the obstruction has been removed, allowing for adjustments in network management. 

This novel set of messages paves the way for enhancing the real-time responsiveness of the network and providing awareness regarding physical obstructions that may impact network performance.

\subsection{5G Standalone Network} 
The OAI 5G Core Network, FlexRIC, the CVF, and the gNB were deployed on an Acer Aspire A715-74G laptop with 16\,GB of RAM and a GeForce GTX 1050 (3\,GB) GPU. The UE was deployed on an HP EliteBook 840 laptop with an 4-core CPU and 8\,GB of RAM. The OAI 5G Core Network was deployed using Docker containers, requiring a 4-core CPU and 16\,GB of RAM~\cite{oai}. FlexRIC is not resource-intensive and has similar requirements to the OAI 5G Core Network.

The gNB and UE were implemented using Ettus USRP B210 Software-Defined Radios (SDRs) due to their cost-effectiveness and popularity within the community. A carrier frequency of 3.6\,GHz was employed for the 5G RAN and two W5084K dipole antennas were attached to each SDR.

\section{Results}
In order to assess the adequacy of the proposed architecture for real-time sensing applications, we first evaluate how the message delay is affected by the number of agents, message size, and message rate. Then, we define a use case testing scenario to assess the functionality of the entire system and implement a vision-aided xApp, which processes the sensing information gathered from the CVF and the SNR values collected from the O-DU via the E2 interface.

\begin{figure}[h!]
    \centering
    \begin{subfigure}{0.45\textwidth}
        \centering
        \includegraphics[width=\textwidth]{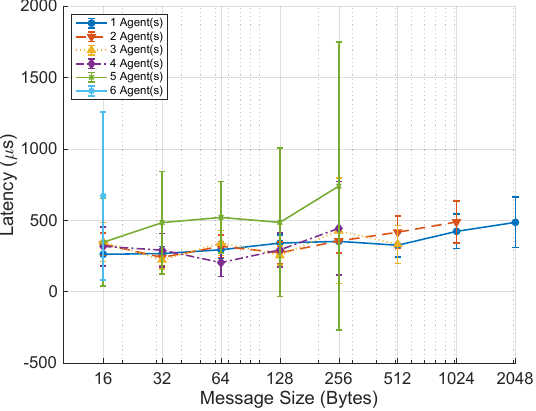}
        \caption{Latency with the increase of message size for different agents (1000 msg/s).}
        \label{fig:CONVERGE_latency_message_size}
    \end{subfigure}
    \hfill
        \begin{subfigure}{0.45\textwidth}
        \centering
        \includegraphics[width=\textwidth]{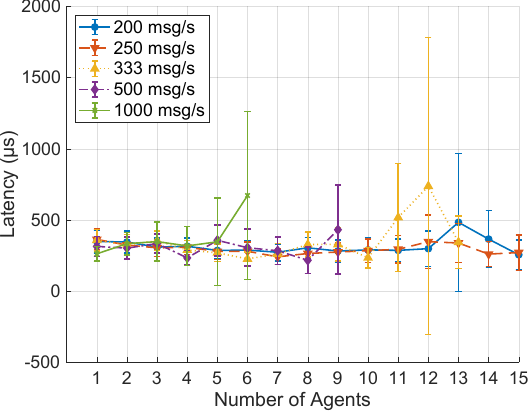}
        \caption{Latency with the increase of agents for different throughput (16 byte message).}
        \label{fig:CONVERGE_latency_agents}
    \end{subfigure}
    \hfill
    \begin{subfigure}{0.45\textwidth}
        \centering
        \includegraphics[width=\textwidth]{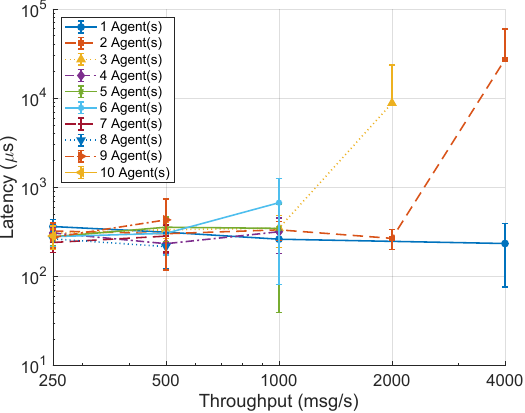}
        \caption{Latency with the increase of throughput for different agents (16 byte message).}
        \label{fig:CONVERGE_latency_throughput}
    \end{subfigure}
    \caption{Multi-agent sensing delay evaluation under different conditions.}
    \label{fig:main}
\end{figure}

\subsection{Message Delay Evaluation}
In the message delay evaluation, we aim to assess the performance of the proposed architecture under different scenarios, measuring the message delay since its generation by an E2' agent outside the gNB until its reception by an xApp. Fig.~\ref{fig:CONVERGE_latency_message_size} shows the latency considering message sizes from 16 to 2048 byte and 1 to 6 agents, generating a rate (throughput) of 1000 sensing message/s per agent. We observe that the average latency is smaller than 1\, ms and that the number of supported agents depends on the message size: 6 agents are supported simultaneously when 16 byte messages are generated, while only 1 agent is supported when 2048 byte messages are generated. If at least 2 agents are required, the sensing messages should not exceed 1024 byte, considering the rate of 1000 message/s per agent. Fig.~\ref{fig:CONVERGE_latency_agents} shows the latency as the number of agents increases under different message throughput, considering 16 byte messages. We observe that the average delay is under 0.5\,ms and that the number of agents depends on the message throughput. For 500 message/s per agent, 9 agents are supported. Fig.~\ref{fig:CONVERGE_latency_throughput} shows the latency with the increase of the message throughput for different number of agents, considering 16 byte messages.  For the selected setup, we can conclude that the number of supported agents decreases with the increase of the throughput of sensing messages, with a limit of 6 agents for one message generated per millisecond per agent, and two agents for a message generated per agent each 200 microsecond. The observed limitations may be caused by the FlexRIC implementation and its system of interrupts required to handle message concurrency, which may have prevented us from testing the usage of more agents under certain message size and message throughput conditions, and may have provided some outliers under heavy load. Improvements in the implementation and more computing power would probably improve the latency values. Even though, the obtained results confirm the support for a relevant number of simultaneous agents introducing low delay, appropriate for the near real-time RAN decisions envisaged for xApps. 

\subsection{Vision-aided xApp and use case validation}
To experimentally validate the combination of radio and video sensing in the RAN under a representative use case testing scenario, we developed a simple vision-aided xApp. It is able to receive and process messages related to environmental conditions, detected by the CVF and made available through FlexRIC, while integrating decoded CVF messages with SNR measurements collected from the E2 interface. The xApp was deployed alongside the FlexRIC to minimize latency between the two components.

\begin{figure}[t]
    \centering
    \includegraphics[width=0.49\textwidth]{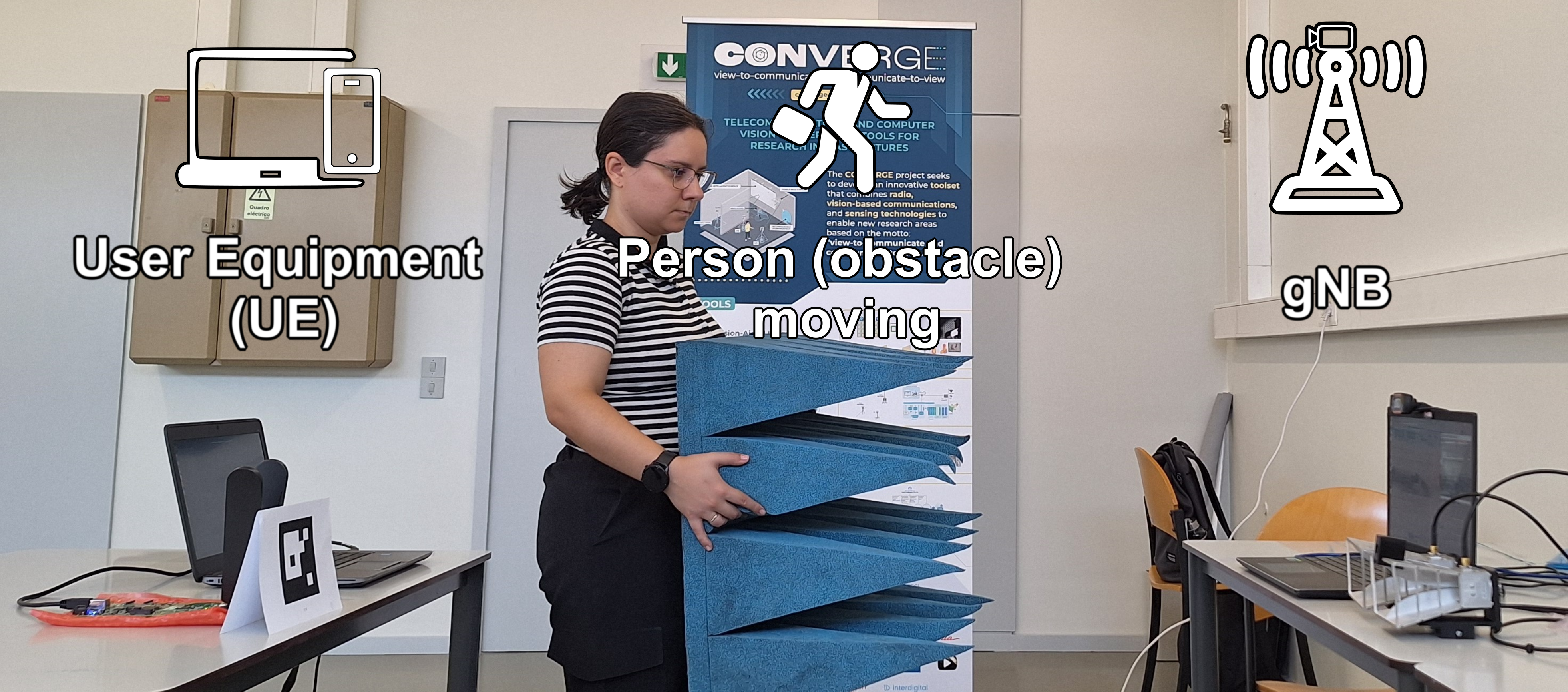}
    \caption{xApp test for the blockage prediction use case including radio and vision data.}
    \label{fig:CONVERGE_combined_figures}
\end{figure}

\begin{figure}[h!]
\centering
\includegraphics[width=0.49\textwidth]{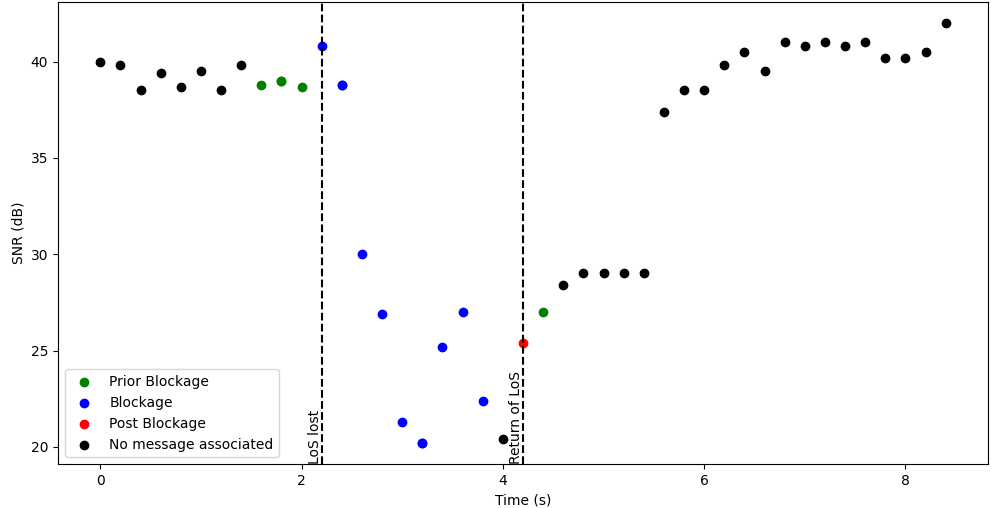}
\caption{SNR of the wireless link between the gNB and UE, along with synchronized CONVERGE Video Function messages, received over time by the vision-aided xApp.}
\label{fig:CONVERGE_xAPP_results_SNR}
\end{figure}

The blockage prediction use case testing scenario was designed to evaluate the impact of blockages on the LoS between the gNB and the UE. By maintaining fixed positions for both the gNB and the UE, we introduced an obstacle to assess its effect on signal quality. This approach also aimed to validate the accuracy of the messages sent by CVF regarding the presence of blockages. The results from this scenario serve as a baseline for evaluating the benefits of integrating CV solutions into 5G and 6G networks.

The experiment demonstrated the vision-aided xApp's ability to receive messages from the CVF along with the associated SNR values for the wireless link established between the gNB and the UE. Fig.~\ref{fig:CONVERGE_xAPP_results_SNR} shows the SNR variation over time, synchronized with each message received from the CVF. SNR values are collected by the vision-aided xApp every 10\,ms, with 20 samples gathered over each 200\,ms interval, aligning with the CVF video processing rate of 5\,fps. With this, each SNR value corresponds to a message received from the CVF.

Fig.~\ref{fig:CONVERGE_xAPP_results_SNR} is divided into three distinct periods, separated by transition lines. In the first period, the obstacle moves towards the UE and \emph{Prior Blockage} messages are received by the xApp. The SNR is high during this period because there is no LoS obstruction between the gNB and the UE. The first transition is labeled \emph{LoS lost}, marking the beginning of a blockage period. In this period, the obstacle stops in front of the UE and the first \emph{Blockage} message is received by the xApp, indicating the beginning of an obstruction. As expected, the average SNR decreases during this period. The second transition is labeled \emph{Return of LoS}, for which a \emph{Post Blockage} message indicates the end of the blockage caused by the obstacle and the return to adequate SNR conditions. In the third period, the obstacle moves away from the UE, no longer blocking the LoS. As expected, the average SNR values are close to those associated with the \emph{Prior Blockage} messages, confirming the removal of the blockage. This graph shows that radio sensing and video sensing provide accurate results by leveraging the combination of multimodal information on the xApp for faster RAN decisions, as the CV algorithm can detect the blockage before it actually occurs, anticipating the decision by 500\,ms.

\section{Conclusions}
The radio channels envisaged for the high frequency 6G radio links will be much affected by the loss of LoS between radio emitters and receivers caused by obstacles that can be easily traced by video cameras. 
In this paper, we present a novel multi-agent architecture that integrates radio and video sensing data to enhance the capabilities of 5G RAN intelligent controllers. By adding CV sensing data to RAN, we enable real-time intelligent controllers such as xApps to make better and faster decisions.

The experimental results obtained validate the effectiveness of the proposed architecture, which was based on O-RAN and FlexRIC, showing that the system can maintain an average sensing delay under 1\, ms even with multiple agents involved. Moreover, the ability of xApps to utilize both radio and video data for network optimization decisions highlights the potential of ISAC in enhancing the Quality of Service (QoS) in next-generation wireless networks. Future work includes closing the sensing and control loop and extending the testbed for outdoor scenarios. 

\bibliographystyle{IEEEtran}
\bibliography{references}

% Generated by IEEEtran.bst, version: 1.14 (2015/08/26)
\begin{thebibliography}{10}
\providecommand{\url}[1]{#1}
\csname url@samestyle\endcsname
\providecommand{\newblock}{\relax}
\providecommand{\bibinfo}[2]{#2}
\providecommand{\BIBentrySTDinterwordspacing}{\spaceskip=0pt\relax}
\providecommand{\BIBentryALTinterwordstretchfactor}{4}
\providecommand{\BIBentryALTinterwordspacing}{\spaceskip=\fontdimen2\font plus
\BIBentryALTinterwordstretchfactor\fontdimen3\font minus \fontdimen4\font\relax}
\providecommand{\BIBforeignlanguage}[2]{{%
\expandafter\ifx\csname l@#1\endcsname\relax
\typeout{** WARNING: IEEEtran.bst: No hyphenation pattern has been}%
\typeout{** loaded for the language `#1'. Using the pattern for}%
\typeout{** the default language instead.}%
\else
\language=\csname l@#1\endcsname
\fi
#2}}
\providecommand{\BIBdecl}{\relax}
\BIBdecl

\bibitem{Saa19}
W.~Saad, M.~Bennis, and M.~Chen, ``A vision of 6{G} wireless systems: Applications, trends, technologies, and open research problems,'' \emph{IEEE network}, vol.~34, no.~3, pp. 134--142, 2019.

\bibitem{Liu22}
F.~Liu, Y.~Cui, C.~Masouros, J.~Xu, T.~X. Han, Y.~C. Eldar, and S.~Buzzi, ``Integrated sensing and communications: Toward dual-functional wireless networks for 6{G} and beyond,'' \emph{IEEE journal on selected areas in communications}, vol.~40, no.~6, pp. 1728--1767, 2022.

\bibitem{EuCNC24}
F.~B. Teixeira, M.~Ricardo, A.~Coelho, H.~P. Oliveira, P.~Viana, N.~Paulino, H.~Fontes, P.~Marques, R.~Campos, and L.~M. Pessoa, ``{CONVERGE}: A vision-radio research infrastructure towards 6{G} and beyond,'' in \emph{2024 EuCNC\&6G Summit}, 2024, pp. 1015--1020.

\bibitem{Charan2021}
G.~Charan, M.~Alrabeiah, and A.~Alkhateeb, ``Vision-aided 6{G} wireless communications: Blockage prediction and proactive handoff,'' \emph{IEEE Tran. on Vehicular Technology}, vol.~70, pp. 10\,193--10\,208, Oct 2021.

\bibitem{Alrabeiah2020}
M.~Alrabeiah, A.~Hredzak, and A.~Alkhateeb, ``Millimeter wave base stations with cameras: Vision-aided beam and blockage prediction,'' in \emph{IEEE 91st VTC2020-Spring}, May 2020, pp. 1--5.

\bibitem{Bonati2022}
L.~Bonati, M.~Polese, S.~D'Oro, S.~Basagni, and T.~Melodia, ``Intelligent closed-loop ran control with xapps in openran gym,'' in \emph{European Wireless 2022; 27th European Wireless Conference}, 2022, pp. 1--6.

\bibitem{Maia2022}
D.~Maia, A.~Coelho, and M.~Ricardo, ``Obstacle-aware on-demand 5g network using a mobile robotic platform,'' in \emph{WiMob 2022}, 2022, pp. 470--473.

\bibitem{Queiros2024}
G.~Queir{\'o}s, P.~Correia, A.~Coelho, and M.~Ricardo, ``Autonomous control and positioning of a mobile radio access node employing the o-ran architecture,'' in \emph{19th Wireless On-Demand Network Systems and Services Conference (WONS)}, 2024, pp. 25--28.

\bibitem{Kim2024}
S.~Kim, Y.~Ahn, D.~Park, and B.~Shim, ``Vomtc: Vision objects for millimeter and terahertz communications,'' \emph{IEEE Transactions on Cognitive Communications and Networking}, pp. 1--1, 2024.

\bibitem{ahn2022toward}
Y.~Ahn, J.~Kim, S.~Kim, K.~Shim, J.~Kim, S.~Kim, and B.~Shim, ``Toward intelligent millimeter and terahertz communication for 6{g}: Computer vision-aided beamforming,'' \emph{IEEE Wireless Communications}, vol.~30, no.~5, pp. 179--186, 2022.

\bibitem{Redmon2018}
A.~Farhadi and J.~Redmon, ``Yolov3: An incremental improvement,'' in \emph{Computer Vision and Pattern Recognition}, vol. 1804.\hskip 1em plus 0.5em minus 0.4em\relax Berlin/Heidelberg, Germany: Springer, 2018.

\bibitem{Alrabeiah2019}
M.~Alrabeiah, A.~Hredzak, Z.~Liu, and A.~Alkhateeb, ``Viwi: A deep learning dataset framework for vision-aided wireless communications,'' in \emph{2020 IEEE 91st Vehicular Technology Conference (VTC2020-Spring)}, 2020, pp. 1--5.

\bibitem{flexric}
M.~Project, ``Flexric: Flexible ran intelligent controller,'' \url{https://gitlab.eurecom.fr/mosaic5g/flexric}, 2024.

\bibitem{schmidt2021flexric}
R.~Schmidt, M.~Irazabal, and N.~Nikaein, ``Flexric: An sdk for next-generation sd-rans,'' in \emph{Proceedings of the 17th International Conference on emerging Networking EXperiments and Technologies}, 2021, pp. 411--425.

\bibitem{bimoflexric}
F.~A. Bimo, R.-G. Cheng, C.-C. Tseng, C.-R. Chiang, C.-H. Huang, and X.-W. Lin, ``Design and implementation of next-generation research platforms,'' in \emph{2023 IEEE Globecom Workshops}, 2023, pp. 1777--1782.

\bibitem{ferreira2023enhancing}
R.~Ferreira, J.~Fonseca, J.~Silva, M.~Tendulkar, P.~Duarte, M.~Ara{\'u}jo, R.~Barbosa, B.~Mendes, and A.~Goes, ``Enhancing network performance based on 5{G} network function and slice load analysis,'' in \emph{2023 IEEE 24th WoWMoM}.\hskip 1em plus 0.5em minus 0.4em\relax IEEE, 2023, pp. 340--342.

\bibitem{Fdi2022}
S.~Fdida, N.~Makris, T.~Korakis, R.~Bruno, A.~Passarella, P.~Andreou, B.~Belter, C.~Crettaz, W.~Dabbous, Y.~Demchenko \emph{et~al.}, ``Slices, a scientific instrument for the networking community,'' \emph{Computer Communications}, vol. 193, pp. 189--203, 2022.

\bibitem{oai}
``Openairinterface | 5{G} software alliance for democratising wireless innovation,'' \url{ https://openairinterface.org//}, accessed on: August 20, 2024.

\bibitem{3gpp}
``3{GPP} ts 38.215, 5{G} nr physical layer measurements,'' \url{https://www.etsi.org/deliver/etsi\_ts/138200\_138299/138215/16.02.00\_60\\/ts\_138215v160200p.pdf}, accessed on: August 20, 2024.

\bibitem{CONVERGE}
``{CONVERGE} project,'' \url{https://converge-project.eu}, accessed on: August 20, 2024.

\bibitem{asn1tools}
E.~Moqvist and Contributors, ``asn1tools: Asn.1 parser and serializer,'' \url{https://github.com/eerimoq/asn1tools}, 2024.

\bibitem{opencv}
I.~Culjak, D.~Abram, T.~Pribanic, H.~Dzapo, and M.~Cifrek, ``A brief introduction to opencv,'' in \emph{2012 Proceedings of the 35th International Convention MIPRO}, 2012, pp. 1725--1730.

\bibitem{Garrido-Jurado2014}
S.~Garrido-Jurado, R.~Muñoz-Salinas, F.~Madrid-Cuevas, and M.~Marín-Jiménez, ``Automatic generation and detection of highly reliable fiducial markers under occlusion,'' \emph{Pattern Recognition}, vol.~47, no.~6, pp. 2280--2292, 2014.

\bibitem{ultralytics_docs}
\BIBentryALTinterwordspacing
U.~Team, ``Ultralytics documentation,'' 2024, accessed on: August 20, 2024. [Online]. Available: \url{https://docs.ultralytics.com}
\BIBentrySTDinterwordspacing

\bibitem{aharon2022bot}
N.~Aharon, R.~Orfaig, and B.-Z. Bobrovsky, ``Bot-sort: Robust associations multi-pedestrian tracking,'' \emph{arXiv preprint arXiv:2206.14651}, 2022.

\end{thebibliography}

\end{document}